\documentclass[a4paper,letters,fleqn,usenatbib]{mnras}

\usepackage{newtxtext,newtxmath}

\usepackage[T1]{fontenc}
\usepackage{ae,aecompl}

\usepackage{graphicx}	
\usepackage{amsmath}	

\usepackage{amssymb}	

\usepackage{rotating}
\usepackage{tikz}

\title[Reproducing the Sagittarius Stream]{Reconstructing the whole 6D properties of the Sagittarius stream with N-body simulations}
\author[Hai-Feng Wang et al.]
{Hai-Feng Wang$^{1}$\thanks{E-mail: haifeng.wang@obspm.fr},
Yan-Bin Yang$^{1}$\thanks{E-mail: yanbin.yang@obspm.fr},
Francois Hammer$^{1}$\thanks{E-mail: francois.hammer@obspm.fr}.
Jian-Ling Wang$^{1}$
\newauthor
\\
$^{1}$GEPI, Observatoire de Paris, Universit\'e PSL, CNRS, Place Jules Janssen 92195, Meudon, France\\
}

\date{Accepted XXX. Received YYY; in original form ZZZ}

\pubyear{2022}

\begin{document}
\label{firstpage}
\pagerange{\pageref{firstpage}--\pageref{lastpage}}
\maketitle

\begin{abstract}

It is a challenge to reproduce the full 6D space-phase properties of Sagittarius (Sgr) dwarf galaxy and its Stream simultaneously. Using N-body simulations with a Milky Way mass of 5.2$\times10^{11}$ M$_{\odot}$ and a ``scaling down''  Sgr mass of 9.3$\times10^{8}$ M$_{\odot}$, from a qualitative point of view, we have been able to reproduce well all 3D spatial features of Sgr stream, including its core, leading and trailing arms, and their associated bifurcations, moreover, the overall trend of the reported 3D kinematics properties of the Sgr stream have also been reproduced without fine tuning. Furthermore, we also find that our model fails in reproducing the exact behaviours of the line-of-sight velocity and angular-energy distributions. It let us to suggest that significant further progress might be achievable after introducing a major component in the Sgr progenitor, which is the gas that dominates all Irregular dwarf galaxies in the Sgr mass range and can slow down the radial velocity of Sgr before its removal, if gas can not solve this problem then we will consider a non-spherical Milky Way halo with hot gas, LMC, etc. As the first step for us to towards the complete understanding of the Sgr system, this progress is also advancing our understanding of the bifurcations, the generation of which might be due to the MW shocks at each pericenter passage, and also be linked to the orientation and disk-shape in the initial conditions.

\end{abstract}



\section{ Introduction }

The Sagittarius dwarf spheroidal galaxy\footnote{We have ultimately decided to defer the modeling of the full Sagittarius dwarf galaxy with gas to a later contribution, which will also be part of the China-France collaboration on the Local group. So please note that this paper II has been delayed to contribute in the future (Paper I: arXiv:2205.02306v3).} is the third most massive system surrounding the Milky Way. It has been discovered by \citet{Ibata1994}, and its associated stellar stream a few years later \citep{Ibata2001,Yanny2000,Newberg2001,Majewski2003,2024ApJ...974..219W}. \citet{Majewski2003} using the 2MASS dataset revealed the presence of two arms of the stream in the south and north sides. With the help of SDSS data-set, precession of the Sagittarius stream and different evolution behaviours of the primary (brighter) tails and the secondary (fainter) tails both in the North and the South were unveiled in \citet{2014MNRAS.437..116B}. The current position and velocity of the Sgr centre are [X, Y, Z] = [17.5, 2.5, $-$6.5] \,kpc, V$_{X,Y,Z}$ = [237.9, $-$24.3, 209.0] km s$^{-1}$ and the Sgr distance from the Sun is around 26.5 \,kpc according to \citet{Vasiliev2020} based on observational data\footnote{X, Y, and Z are in the left-handed Galactocentric cartesian coordinate same as this work, the sun location is at the left side of the Galactic centre.}.

 \citet{Ibata2020} has given the six-dimensional panoramic portrait of the Sagittarius stream, based on the Gaia DR2 dataset and  the STREAMFINDER algorithm. They found that the global morphological and kinematic properties of the Sagittarius stream could be matched well with the model made by \citet{Law2010}. With the help of LAMOST M giant stars, \citet{Lijing2019} used 164 pure Sgr stream stars to find that the trailing arm has higher energy than the leading arm at given angular momentum (see also \citealt{Vasiliev2021} and \citealt{Hammer2021}). For more details about Sgr stream, the north bifurcation was discovered in \citet{Belokurov2006} and south bifurcation was discovered in \citet{2012ApJ...750...80K}.
Recently, \citet{Ramos2021} also successfully detected two bifurcations, one in each of the northern leading arm and southern trailing arm, which implies that two bright and two faint branches are needed to fully reproduce the Sgr stream. The four branches in \citet{Ramos2021} are much clearer thanks to the precise Gaia EDR3 dataset. Following this \citet{2022ApJ...932L..14O} have also recovered half of the branch, the faint bifurcation in kinematics and also star counts, using the N-body simulations GYRFALCON \citep{Dehnen2000} based on the \citealt{Vasiliev2021} by adding test particles on disk initial orbits of the Sgr.

Most models cannot reproduce the stream bifurcations (see  Fig. 4 of \citealt{Ramos2021}), including the model of \citet{Law2010} that matches well the Gaia DR2 data, and the more recent model by \citet{Vasiliev2021}. \citet{Vasiliev2021} suggested that the model with a massive LMC can match well misalignment between stream track and sky-plane velocity in the Sgr coordinate, suggesting us that it might require a LMC mass of about 1.3 $\times10^{11}$ M$_{\odot}$ associated to a Milky Way mass near 5.6$\times10^{11}$ M$_{\odot}$ within 100 \,kpc and virial  mass is about 9$\times10^{11}$ M$_{\odot}$. The observed bifurcation is not fully retrieved and without discussing gas physics influence as mentioned in \citet{Vasiliev2021}. Moreover, they used two models with/without LMC to reproduce the Sgr core and stream, perhaps it is reasonable to ask for us now whether one model without physical changes but considering the gas contribution can solve the misalignment problems and leading arm overestimation issues. 

So starting from this work, we want to explore more about this intriguing topic related to the Sgr and gas in the Milky Way and dwarfs, but before that we need to reproduce the 6D properties with pure N-body simulations in the same model. As far as we know, the only model able to reproduce the bifurcation in the leading tail of the Sgr stream was made by \citet{Penarrubia2010} after modeling the initial Sgr as a late-type, rotating disc galaxy, then \citet{2011ApJ...727L...2P} conducted spectroscopic surveys to show evidence that models with little or no intrinsic rotation is better matched to the mean line-of-sight velocity, but fail to reproduce the shape of the line-of-sight velocity distribution. However, at that time, the model could not be kinematically constrained in absence of Gaia proper motions, and we also notice that it is not surprising that the initial model galaxies are exponential disks embedded in the spherical dark matter halos for Local Group dwarfs, which were investigated quite well in \citet{Mayer2001, 2015ApJ...810..100L}. More recently, \citet{2022ApJ...935...14C} also confirmed the importance of ro- tation of Sgr dwarf galaxy using N-body simulations in order to investigate more about the M54 globular cluster embedded in a dark matter halo.

As known to the community, despite many modeling efforts have been done to reproduce the Sgr stream properties (e.g., \citep{Ibata2001,2001MNRAS.323..529H,Helmi2004,Martinez-Delgado2003,Law2005,Law2009,Law2010,Fellhauer2006,Penarrubia2010,2020ApJ...889...63H,Vasiliev2021}, almost none of group can perfectly explain or reproduce the wealth of observations, especially the bifurcations, which masks this topic is more and more fascinating in the Gaia era.

In this work, motivated by the recent Gaia observational clearer features and some modelling implications, we attempt to use N-body modeling method to explore more about 3D shape of the stream, core, bifurcations, leading and trailing arm in order to better understand the full features of the Sgr stream. This paper is structured as follows, in Section 2, we focus on the modeling details of our simulation, in Section 3 we discuss the observational properties to be reproduced, and in Section 4 shows the comparison between observations and modeling. The last Section includes the discussion and then the conclusion. This work plus \citet{wang2022} perhaps can be considered as the first step for us to towards the better understanding for the well-known Sgr system in our group.
\section{N-body simulation and model}\label{method}
Our Milky Way model is composed of a shallow cusp bulge \citep{Hernquist1990}, an exponential disc \citep{Freeman1970} and a cored dark matter halo \citep{Dehnen1993}, as described by the following density profiles:

\begin{equation}\label{eq:thick_nu1}
\rho_{\text{bulge}} \propto r^{-1}(r + a_{\text{bulge}})^{-3}
\end{equation}

\begin{equation}\label{eq:thick_nu2}
\rho_{\text{disk}} \propto \exp\left(-\frac{R}{R_{\text{disk}}}\right) \, \text{sech}^2\left(\frac{z}{z_{\text{disk}}}\right)
\end{equation}

\begin{equation}\label{eq:thick_nu3}
\rho_{\text{halo}} \propto (r + a_{\text{halo}})^{-4}
\end{equation}

The model of Sgr Dwarf spheroid galaxy is made of a dark matter halo and a disc which are described by Eq.\ 2 and Eq.\ 3 respectively. 
 
Simulations in this paper are carried out by using GIZMO \citep{Hopkins2015}. N-body realisation of initial conditions are created using Zeno\footnote{\url{https://home.ifa.hawaii.edu/users/barnes/software.html}}. In our N-body simulations, the Milky Way particle mass resolution is 1.375$\times10^{5}$ M$_{\odot}$ with 1.2  Million particles and Sgr mass resolution is 1.375$\times10$ M$_{\odot}$ with 18 Million particles. 

To accommodate the large mass ratio between dark matter particle and stellar particle (Table~\ref{table1}), we adopted the Adaptive Gravity Softening (AGS) in GIZMO including Lagrangian conservation terms \citep[see][]{Hopkins2018,Hopkins2015}. The AGS technique has been approved to be efficient to eliminate the possible artificial kicks between dark matter particle and stellar particle in \citet{wang2022}. Thus we adopted the same method as in \citet{wang2022} here and study the whole streams quantitatively and qualitatively for the basic 6D properties.

Table~\ref{table1} details the initial conditions of our dynamical simulations. We note that the mass adopted for the Milky Way is consistent with its rotation curve as it is included in the \citet{Jiao2021,2023A&A...678A.208J} range, and is very close to that recently found by \citet{Wang2022}, both of which mass range are confirmed indirectly by recent Gaia DR3 disk rotation curve work with the slope of $-$2.3 $\pm$ 0.2 km s$^{-1}$ kpc$^{-1}$ \citep{wang2023a,2023arXiv230900048J}. The adopted mass of the Milky Way here is considerably lower than most current estimates \citep{2020SCPMA..6309801W}.

The initial Sgr starts to orbit the Milky Way from its initial location following an orbit near polar, and in a prograde-retrograde mode \citep[see the detailed definition in \citealt{Sauvaget2018}]{2002MNRAS.333..481B}. We have then used different snapshots at different epoch to compare the results with observations, and finally we  find that 4.7 \,Gyr is the best epoch, which allows a third passage. 

\begin{figure*}
  \centering
  \includegraphics[width=1.0\textwidth]{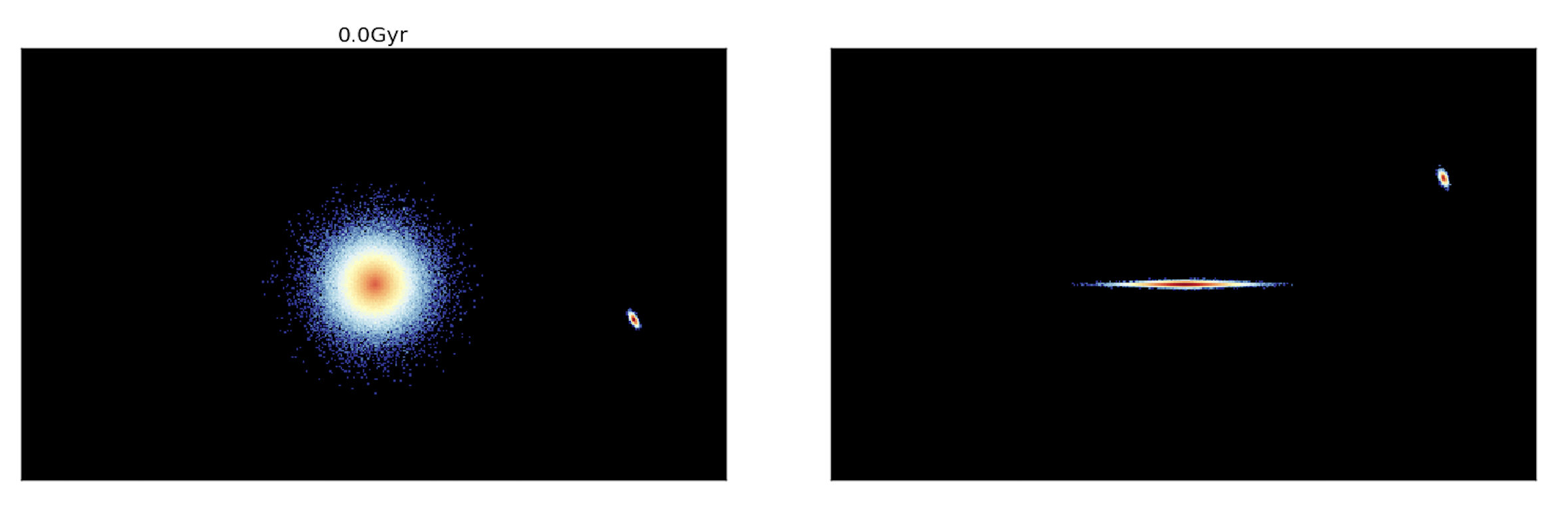}
  \caption{It shows an overview of the Sgr orbit in the face-on and edge-on plane for our initial conditions, the exact initial value is shown in Table~\ref{table1}.}
  \label{SgrmodelIC}
\end{figure*}

\begin{figure*}
  \centering
  \includegraphics[width=0.8\textwidth]{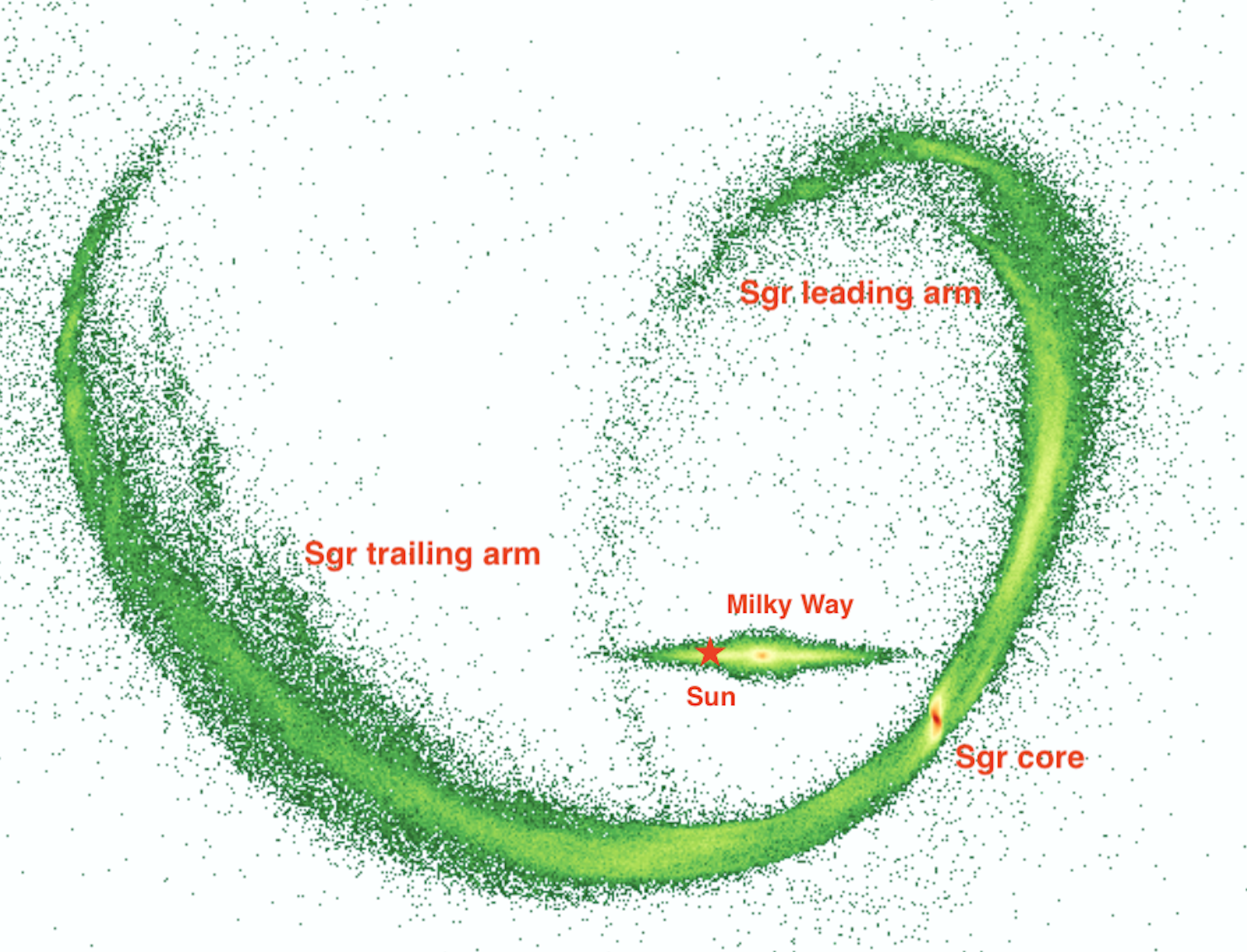}
  \caption{It shows an overview of the Sgr orbit in the orbital plane, where the Milky Way is seen edge-on. This simulation result corresponds to our present day modeling in many hundreds of simulations, showing clearly the trailing arm, the leading arm, the core, and the possible two bifurcations, all labeled in red. Milky way disc and the Sun's location are also shown. The north bifurcation is clear as mentioned in many models, the most likely south bifurcation signal is also shown here. More comparisons could also be found in Figure~\ref{3Dcomparison}.}
  \label{Sgrmodel}
\end{figure*}

The initial snapshot in the face-on and edge-on plane is shown in Figure~ \ref{SgrmodelIC}, and Figure~ \ref{Sgrmodel} shows our modelling result corresponding to the present day state, near the orbital plane in which the Milky Way is edge-on, after 4.7 \,Gyr. We could see clearly the core, leading arm, trailing arm, and the bifurcation signals (edge-on) in the north side found in many literature and a less prominent southern bifurcation signal revealed by \citet{Ramos2021}. Many models also have the similar spatial patterns, so we will have some more investigations in the next part about the bifurcation signals.

\begin{table*}
\centering
\caption{Initial condition parameters in this dynamical simulation.} 
\label{table1}
\centering
\begin{tabular}{lllll}
\hline
Parameter
         &
Milky Way    &
Sagittarius        &
Units             \\
\hline
Particle mass (star/dark matter)         &1.375(5.575)$\times10^{5}$  & 1.375(5.513)$\times10$ & M$_{\odot}$\\
Dark Matter mass     & 4.8$\times10^{11}$  & 9.0$\times10^{8}$  & M$_{\odot}$\\
Dark Matter scale     & 9.1 &1.6 & kpc\\
Stellar disc mass  & 3.58$\times10^{10}$    &   3.0$\times10^{7} $   & M$_{\odot}$\\
Stellar scale length & 2.4    & 0.3   & kpc\\
Stellar scale height & 0.24   & 0.15 & kpc\\
Bulge mass& 1.12$\times10^{10} $   & N/A  & M$_{\odot}$\\
Bulge scale & 0.4   & N/A   & kpc\\
Number of particle & 1.2  & 0.18    & Million\\

Spin & (l, b)=[108, -90 ]  & (l, b)=[30.1, 20.2 ]    & degree\\
\hline
Orbit pole (MW+Sgr) &\qquad \qquad   (l, b)=[268, -15.6]    degree\\
\hline
Initial Position &   &(66, -9, 27)& \,kpc \\
\hline
Initial Velocity &   &(-48, -17, 65)& \,km s$^{-1}$  \\
\hline
\end{tabular}
\end{table*}

\section{Observational Data} \label{Observational Data}

In this work, we have made full use of the Red Giant Branch (RGB) catalogue with 55 000 stream members from \citet{Vasiliev2021}, including  about 4500 stars with line-of-sight velocities from various spectroscopic surveys. Star distances have been approximated by the median distance of the five nearest RR Lyrae stars on the sky, and the extinction from \citet{Schlegel1998}. RR Lyrae stars are from Gaia DR2 \citep{Gaia2018a} and are listed into two catalogues from \citet{Holl2018} and \citet{Clementini 2019}. Additionally, we have used the 44,000 RR Lyrae (RRab) stars without line-of-sight velocity from the Pan-STARRS1 (PS1) 3$\pi$ survey provided by \citet{Hernitschek2017}.

$\Lambda$ and $B$ are the longitude and latitude in the Sgr stream coordinate system very similar to \citet{Majewski2003}. In this work, the leading arm is from 60$^{\circ}$ to 180$^{\circ}$, the trailing arm covers a range of 180$^{\circ}$ to 360$^{\circ}$, small differences for this will not change the conclusion here. For the Milky Way coordinates in this work, the core/stream is projected in a left-handed Galactocentric cartesian coordinates [X, Y, and Z]. We have adopted the solar motion from \citet{2018RNAAS...2..210D} with [$U_{\odot}$ $V_{\odot}$ $W_{\odot}$] = [12.9, 12.6, 7.78] km s$^{-1} $. The circular speed of the Local Standard of Rest is adopted to be 233 km s$^{-1} $ \citep{2018RNAAS...2..210D}. The distance of the Sun from the Galactic center is chosen to be $R_{\odot}$  = 8.277 \,kpc \citep{2022A&A...657L..12G} and $Z_{\odot}$ = 20.8 \,pc \citep{2019MNRAS.482.1417B}, and we have verified that different solar motions and coordinates would not change our final conclusions, either. 

\section{Results} 

\begin{figure*}
  \centering
  \includegraphics[width=1.0\textwidth]{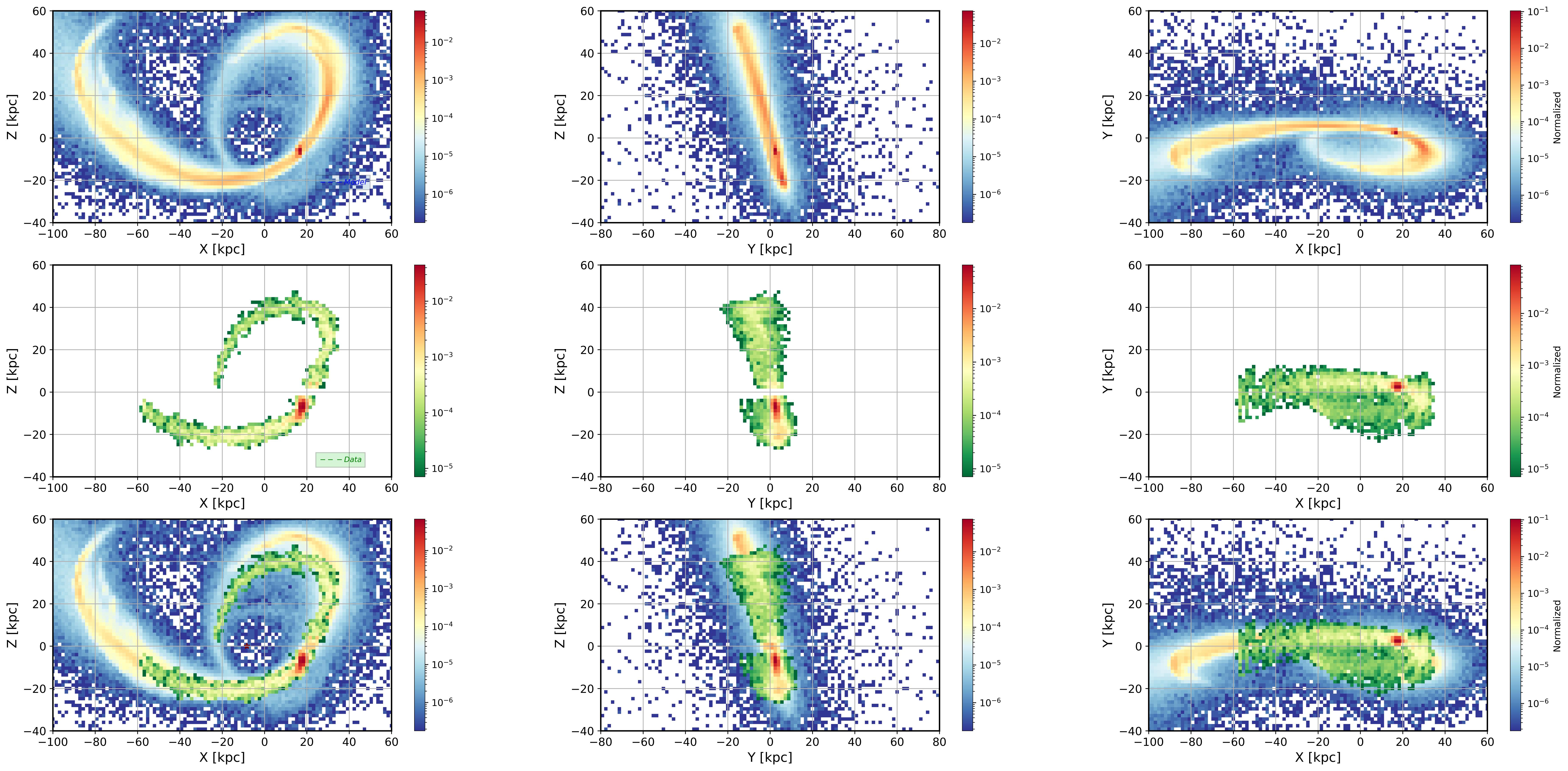}
  \caption{The top panel shows the simulated model, the middle panel shows the observational stream from the RGB stars, and the bottom panel is the superposition of the modeled particles with the observations. From left to right, the distribution in the X-Z plane, Y-Z plane and X-Y plane are shown, respectively. The colour code reflects the number counts (see bars on the right of each panel).}
  \label{3Dcomparison}
\end{figure*}

Figure~ \ref{3Dcomparison} presents a detailed comparison of our model with observational data. It shows the projection of the stream in 3 planes (XY, YZ, XZ), X, Y, and Z are in the conventional Galactocentric cartesian coordinate. Top, middle, and bottom panels show the modeled stars, the observed RGB star distribution, and the overlap of the data and model, respectively. In all panels the core location is identified by the density peak value (red colour), similarly as in  \citet{Vasiliev2021} for the location, and at about [X, Y, Z] = [17.5, 2.5, $-$6.5] \,kpc. The size is relatively smaller due to that we start from a ``scaling down''  Sgr progenitor so our velocity dispersion is a bit smaller by few km s$^{-1}$ than \citet{Vasiliev2020}. It is can be noticed that the trailing arm in our model is more elongated than that shown by the \citet{Vasiliev2021} RGB stars. Moreover, it can be seen that we have reconstructed very clear north bifurcation at the top and bottom left panel, which is shown around X = 30\,kpc, and some south bifurcation signal might be also shown around X = $-$60\,kpc. Note that it starts to be bifurcated from around $-$20 to $-$30 \,kpc, which is not the same as previous works in the conmmunity, except the bifurcation sigals, we have no other better explanation for the moment.

\begin{figure*}
  \centering
  \includegraphics[width=0.8\textwidth]{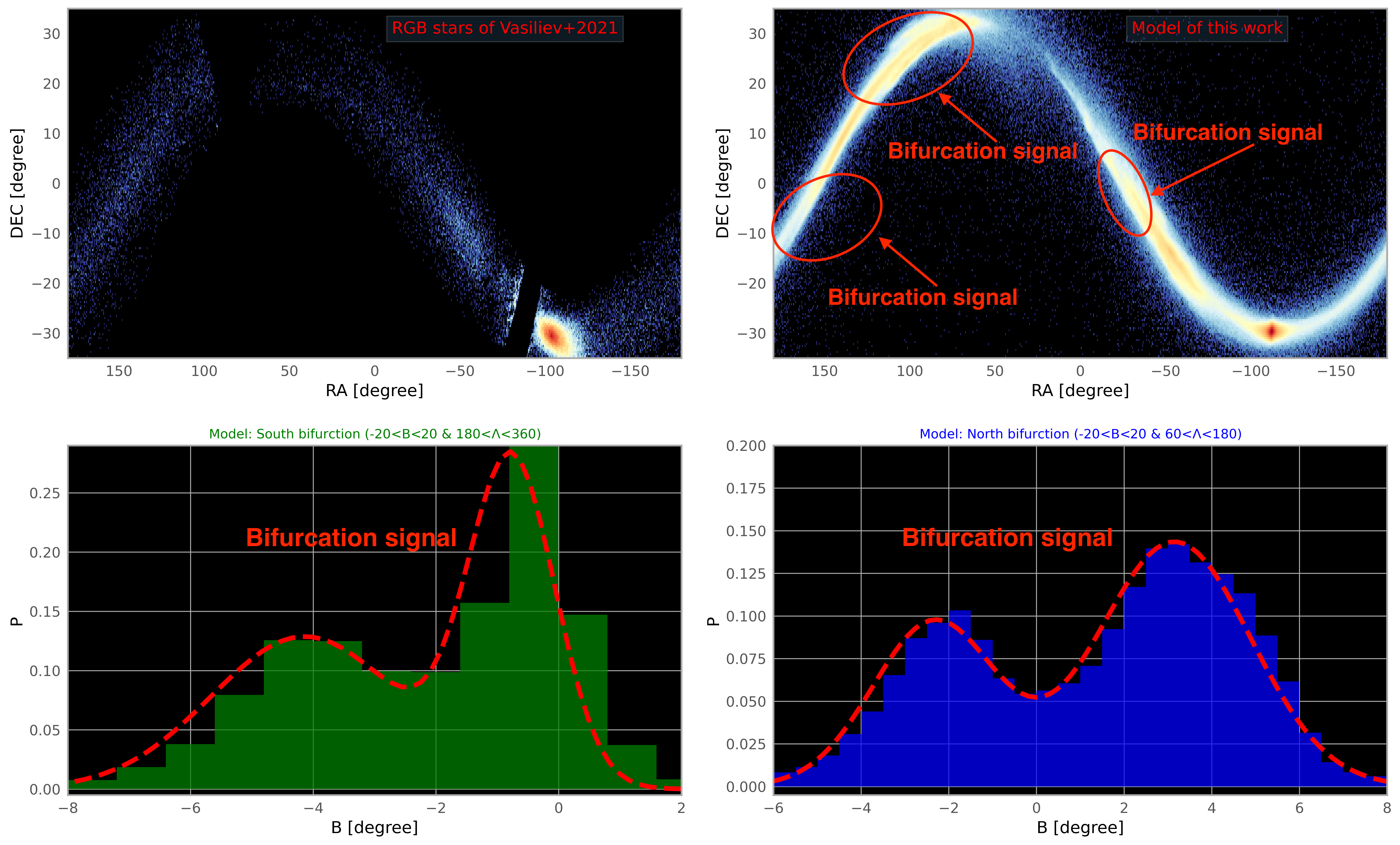}
  \caption{Our model for bifurcation signals in two coordinate systems which can be easily converted to each other: top left one shows the distribution $RA$ vs. $DEC$ in Galactic equatorial coordinate and top right is $\Lambda$ vs. $B$ in Sgr coordinate and some bifurcations signals are clearly marked (one can zoom a bit it). The double gaussian fitting results are showing here in the bottom two panels by red dashed lines, which are corrsponding to the trailing/leading faint and bright branches, the range we choose to find bifurcation signals are shown as the title in the bottom two subfigures, these figures here can be compared with recent bifurcation progress of Fig. 2 of \citet{Ramos2021} but the exact match is not expected for the moment, we only focus on the signals not the full reconstruction, such like fractions of different branches. $\Lambda$ and $B$ are the longitude and latitude in the Sgr stream coordinate system very similar to \citet{Majewski2003}, the leading arm is from 60$^{\circ}$ to 180$^{\circ}$, the trailing arm covers a range of 180$^{\circ}$ to 360$^{\circ}$ for the $\Lambda$ and $B$ is within 20$^{\circ}$. Here we only focus on the bifurcation signals and we defer the exact analysis of the line of sight velocity to the future work.}
  \label{radecsgr}
\end{figure*}

Figure~ \ref{radecsgr} shows our modeling in the euatorial or Sgr coordinates, i.e., RA vs. DEC or $\Lambda$ vs. $B$ plane, the RGB patterns is clearly shown in the top left panel in the equatorial coordinates, which is reconstructed well on the top right panel. One can compare it to the Fig. 2 \citet{Ramos2021} results qualitatively. As delineating in observations, there are some bifurcation signals in the modeling, which are hidden in the top right panel marked with red arrows and labels. The bifurcation is better seen in the two bottom panels that show the particle distribution along the Sgr $B$ declination, which can be regarded as made of two gaussians naturally in both south and north sides (see red dashed lines), to which the particle distribution is superimposed. Our modeling is then consistent with the \citet{Ramos2021} findings, i.e., both observations and modeling show the bifurcation signals for each Sgr stream arm, which supports that initially Sgr possessed a rotating disk. Notice also that \citet{delPino2021} analyzed the full 6D structure of the Sgr core, and found an evidence for a residual, though moderate rotation. And meanwhile, the analysis of the \citet{Ramos2021} shows that all current models can not fit the Gaia EDR3 results perfectly and rotating progenitor are worth exploring.

We have qualitatively compared Figure~ \ref{radecsgr} in this work with Figure 2 of \citet{Ramos2021} and we notice that the bifurcation signals observed in this model are not the exact the same but not controversial. Since this model is still ongoing, perhaps we don't expect to reconstruct the very exact fraction as shown in the \citet{Ramos2021} EDR3 work.

\begin{figure}
  \centering
  \includegraphics[width=0.45\textwidth]{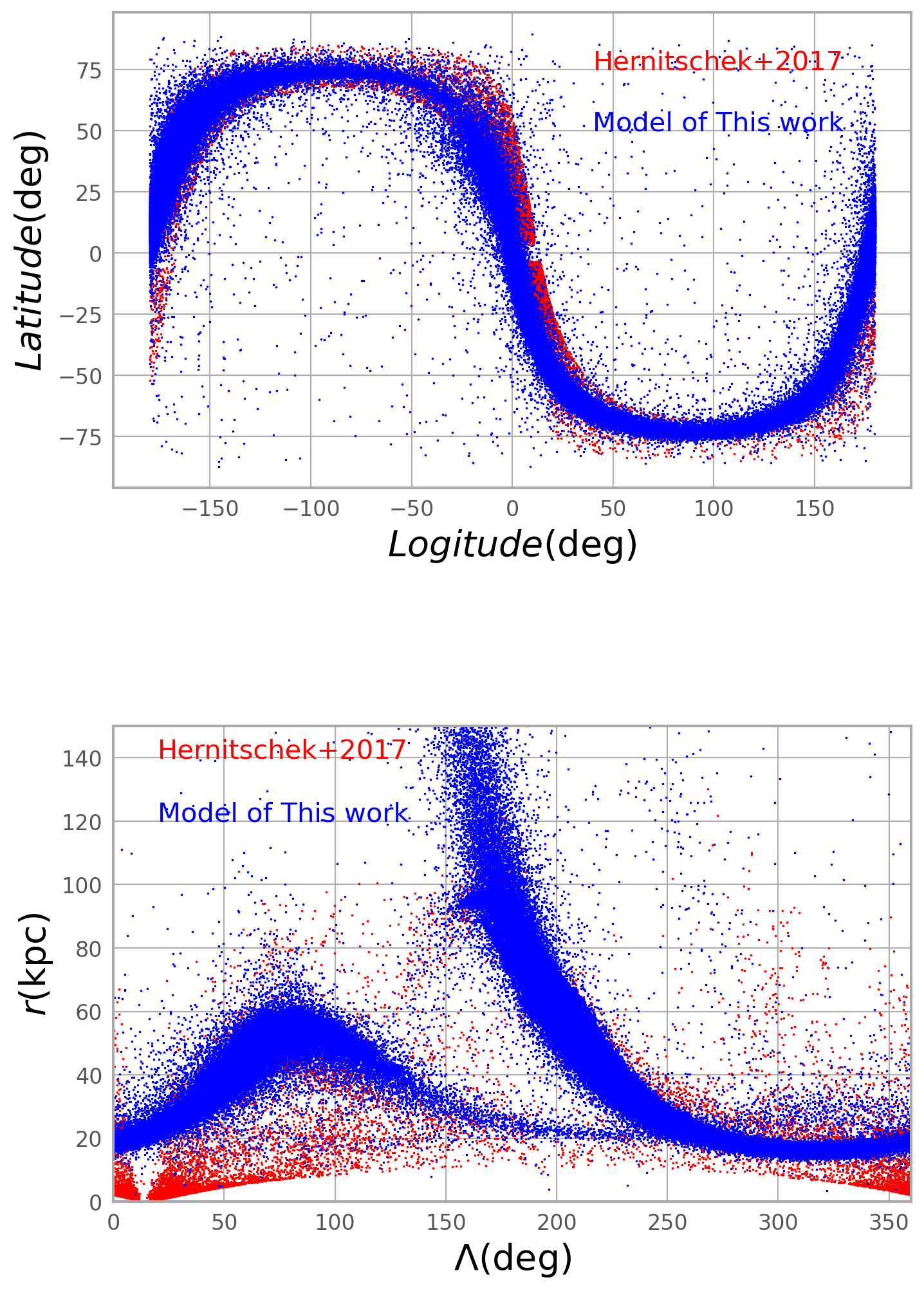}
  \caption{The figure compares RR Lyrae stars (yellow points) with our model (blue points) in the longitude vs. latitude plane (on the top panel) and  Galactic distance vs. $\Lambda$ plane on the bottom panel, which is matched quite well.} 
  \label{vb2021PMgra}
\end{figure}

The panel of Figure~ \ref{vb2021PMgra} shows the comparison between our model and RR Lyrae stars (yellow dots) from \citet{Hernitschek2017}. It is clearly showing a good match for both the leading and trailing arm. As seen, for the 3D spatial distributions, this model can reproduce the observational features well.  

\begin{figure}
  \centering
  \includegraphics[width=0.45\textwidth]{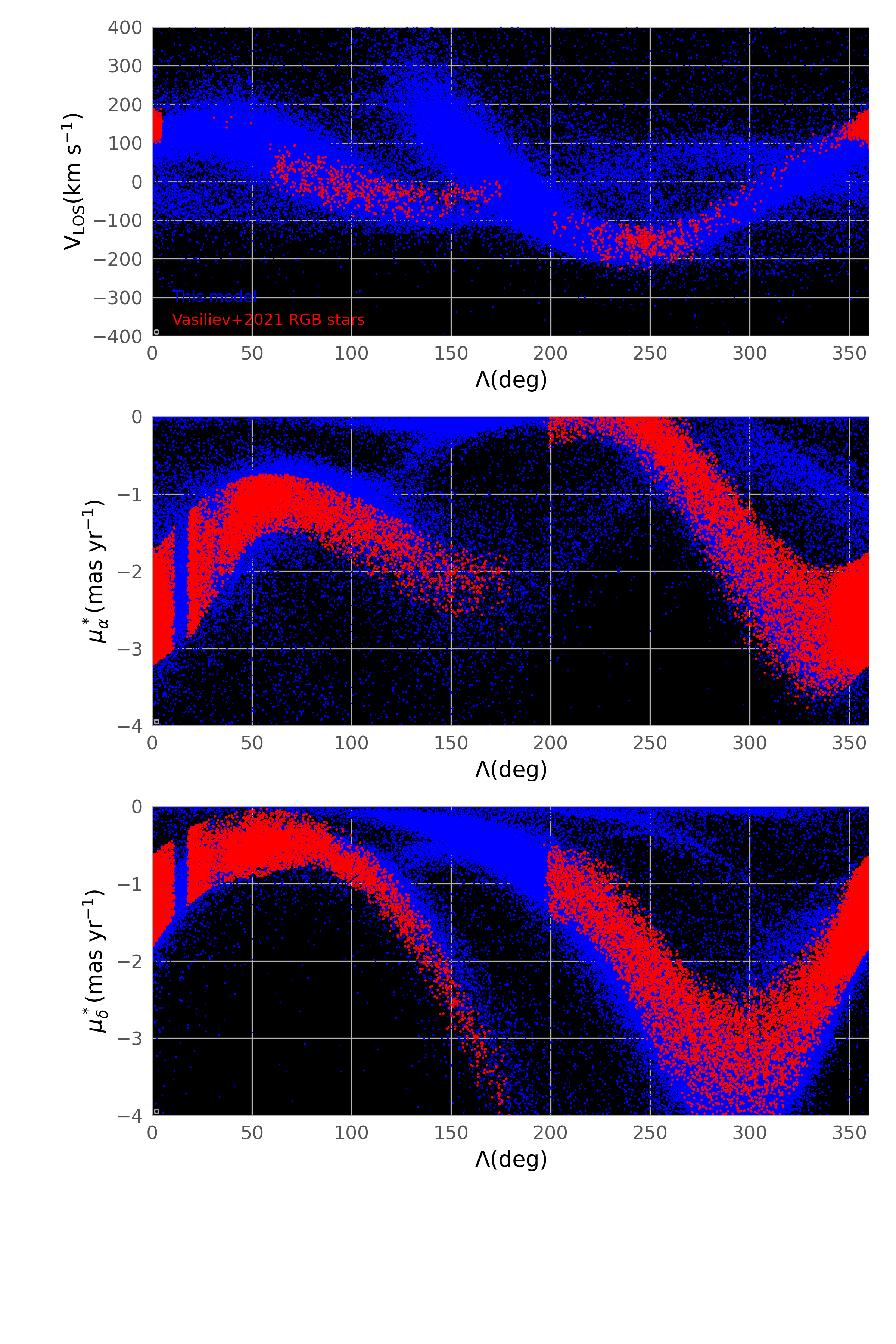}
  \caption{The top panel shows line of sight velocity along with the $\Lambda$ Sgr coordinate, for the model (blue dots) and for RGB stars (red dots) with line-of-sight velocity. The bottom two panels provide the two proper motions along the $\Lambda$ axis for model and RGB star data. Note the general pattern is matched quite well for proper motion. Some offsets can be found in the top panel, which will be solved by more fine tuning process and detailed physical process in the next work.} 
  \label{vb2021PMgraa2}
\end{figure}

Then we present the 3D kinematics analysis in Figure~ \ref{vb2021PMgraa2}, the top panel of Figure~ \ref{vb2021PMgraa2} presents the variation of the line-of-sight and proper motions along the $\Lambda$ axis, showing that model particles (blue dots) almost follow consistently RGB stars (red dots) for the overall trend, some offsets are appeared, for example, the line-of sight velocity around 0 degree. The proper motion is much better than the radial velocity with very small offset.

The qualitative match, as seen, can be acceptable at this stage, some clear differences might be due to the fact that data cannot be homogeneously distributed on the sky because of the Milky Way disk contamination, and our model still not perfect, for example we ignore gas which is known to not be easily considered. In short, from the qualitative point of view, we reconstruct the 3D kinematics patterns of the Sgr stream and in the meantime, the core can also be reproduced in the exactly same one model (see also \citet{wang2022}).  

Frankly, as far as we know it is unclear whether simply more fine tuning is necessary for improving the modeling, or alternatively if the model needs supplementary ingredients, such as an intial Sgr dominated by gas, or a massive LMC as in \citet{Vasiliev2021}, more works will be shown in the future.

For the gas argument or framework, recently, \citet{2023MNRAS.520.1704B} have already predicted the infall time of the Sgr is about 8 Gyr and the quenching time is about 5 Gyr, which has also clearly encouraged us that it is non-trival to consider gas and star formation history. In addition, \citet{Thor2018}, with the simulations for Sgr gas, also have showed that the gas stripping was 30$-$50 percent complete at its first disc crossing around 2.7 Gyr ago, then entirely stripped at its last disc crossing 1 Gyr ago, but they didn't compare more in details about the 6D properties.

\begin{figure}
  \centering
  \includegraphics[width=0.45\textwidth]{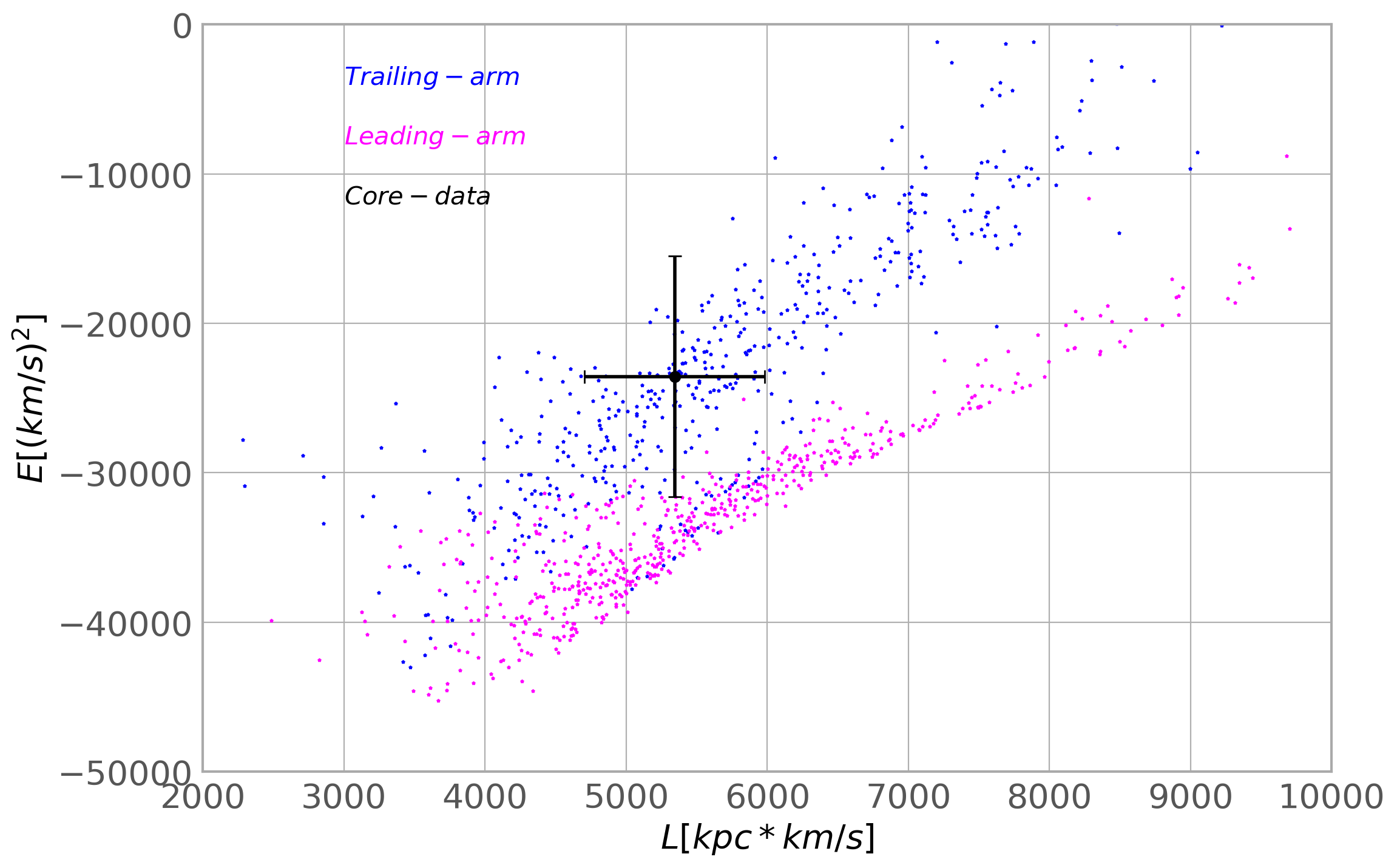}
   \centering
  \includegraphics[width=0.45\textwidth]{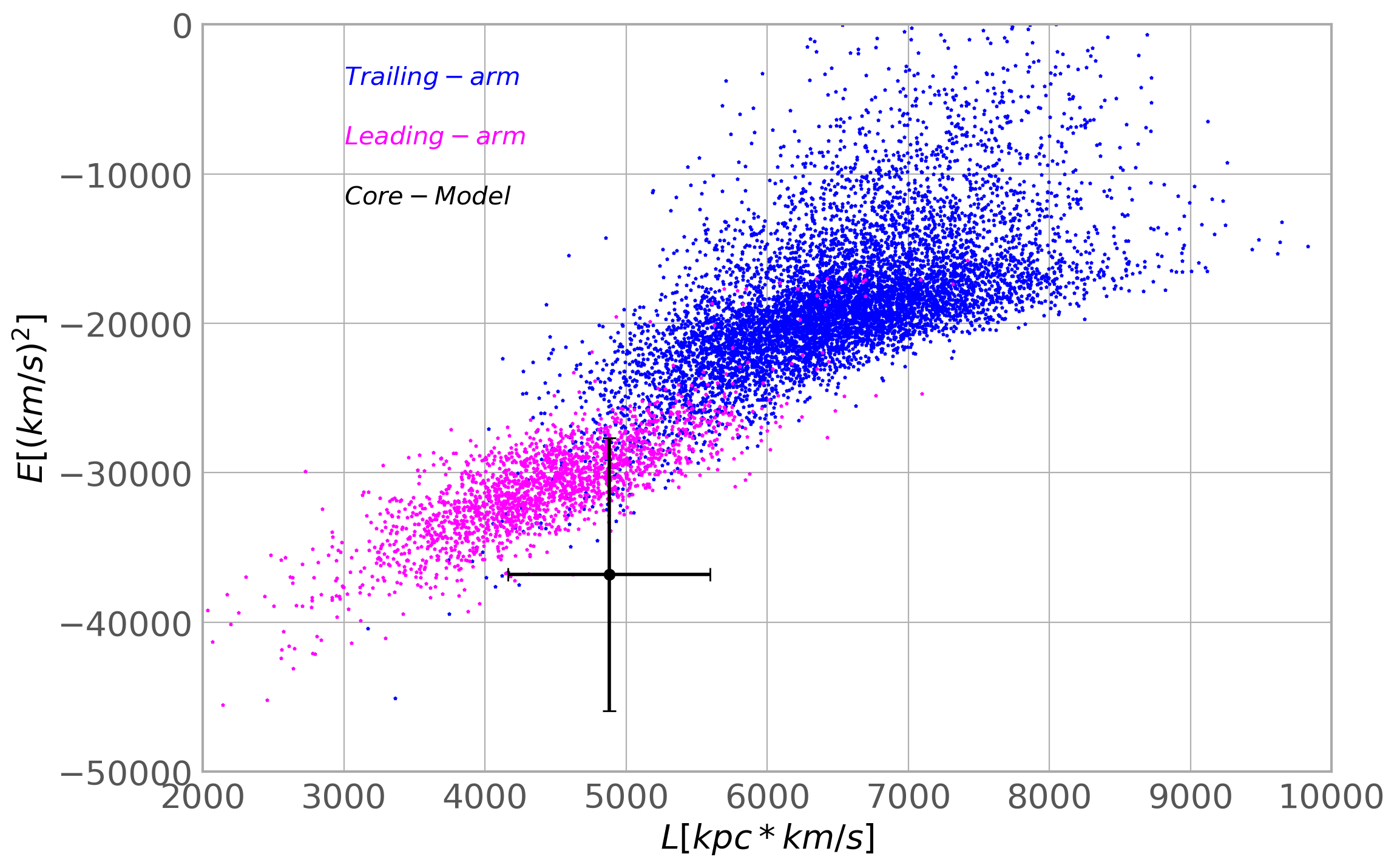}
  \caption{The Figure shows the behaviors of both leading (magenta dots) and trailing (blue dots) arms selected from the Sgr coordinates ($\Lambda$, $B$). The top and bottom panel shows the RGB stars, and model particles in the energy-angular momentum plane, and the black cross identifies the Sgr core location and error bars.}
  \label{EL1}
\end{figure}

Figure~\ref{EL1} shows the angular momentum ($L$) and energy ($E$) distribution for the trailing arm (blue dots) and leading arm (magenta dots), for RGB stars with line of sight velocities (top panel) and for modeled particles (bottom panel). Here we use the exponential disc potential, core model potential for halo and bulge mentioned in Section 2 to calculate energy for RGB stars and particles. The scarcity of the data points in the top panel is due to the limited number of RGB stars that have been spectroscopically observed. Here we find that the model does not reproduce the data exactly enough. Actually data shows that both arms with a different behavior in the (E, L) plane that can be approximated by an almost linear relation extending from L= 3000 to L= 10000 kpc$\times$km$s^{-1}$. As seen, the model also predicts a more energetic trailing arm, which is similar to the observations, but we find that both arms likely to follow the similar (E, L) relation. This is the main indication of a disagreement between modelling and observations, which requires further discussion (see next section).

\section{Discussion}  \label{Concl}

\subsection{Model comparisons}  

In order to better understand the Sgr system, we firstly have a brief summary of the recent modelling progress from \citet{Vasiliev2020, Vasiliev2021}. By focusing on the Sgr remnant, \citet{Vasiliev2020} suggested that they can not fit the stream well in the same model, in particular for the leading arm, and the disky model is worse for the remnant reconstruction. During that same work, they used the spherical less-massive MW potential with particular choice and spherical non-rotation Sgr model without LMC and dynamical friction. Following this, \citet{Vasiliev2021} adopted a second model with MW + LMC suggesting that it explains the misalignment between stream track and sky-plane velocity. 

In our new model we don't consider the LMC for the moment, which is not the main target at this stage, but perhaps we can attempt to propose a new possibility, that is to say, the misalignment might be caused by the gas (energy), which will be investigated more in our series of works. Other possibilities such like rotated milky way halo potential for the misalignment was also proposed recently \citep{2021ApJ...910..150V} so we think it is non-trival here to propose this new possibility that the misaligment might be solved by considering the cold gas of the Sgr, hot and cold gas of the Milky Way \citep{2009ARA&A..47...27K,2002MNRAS.333..481B}, time-dependent non-symmetrical potential in the future. 

Moreover, \citet{Vasiliev2021} in their modeling paper of the stream claimed that a spherical halo is not good at fitting for the morphology of the stream (especially leading with careful reflex corrections), but the model in this work can fit it at least for the overall trend. They also claimed that without LMC, the distance pattern is worse, it seems that it is not unacceptable here using spherical halo and cored model to reveal the north bifurcation, although more tests in our modelling will be needed. As one of the success, their halo profiles are very flexible including static/frozen/restricted/live choices, and more importantly, dynamical friction is considered in the stream modelling. The initial time in their model is from 2.5-3 Gyr ago with massive MW and with/o massive LMC. 

Interestingly, a recent work finished by \citet{Koposov2023} shows that the LMC has a significant influence on the Orphan-Chenab large stream (OC), and the prolate and oblate halo is better than the spherical halo. However, OC stream is not consistent with Sgr stream for the potential shape. They can not fit the energy and angular momentum perfectly either, and they also thought their model is not perfect and a lot of works needed. So it implies that each stream would need one specific LMC and MW, which makes this topic more complicated but interesting.  

In short, using non-spherical MW halo and NFW spherical LMC and spherical non-disky Sgr, \citet{Vasiliev2020, Vasiliev2021} established two new models about the Sgr system and they also pointed that the residual rotation might be better, bifurcation was failed and encourage our community for more models including MW and Sgr gas. More recently, \citet{2022ApJ...932L..14O} revisited a disky origin for the faint branch of the Sagittarius stellar stream by adding test particles based on the  \citet{Vasiliev2021} N-body simulation, and the infall time is around 3 Gyr, which is different than current work for the low mass potential and initial infall time, they can almost recover half of the bifurcations including the self-gravity of this disky component, and they also analyzed the line-of sight velocity showing the good match for the faint branch, and there are some clear offset in the bright branch, so there are remaining a lot of mysteries of the bifurcation origins and the progenitor of the Sgr.

So thanks to the \citet{Vasiliev2020, Vasiliev2021} nice progress, we have made use of their initial orbits information but by scaling down their orbits to match our relatively lower Milky Way-Sgr mass system. But we have some different theoritical possibilities, so this model might be seen as a new attempt after the progress of the \citet{Vasiliev2020, Vasiliev2021}, although it is still ongoing, the promising modelling might have been shown. 

More importantly, we would like to point out that all existing models of the Sgr stream are lacking of gas, which could have interacted with the Milky Way hot gas through ram-pressure stripping as these effects are preponderant for the Magellanic Clouds and Stream \citep{Hammer2015,Wang2019}. We tentatively associate the main failure of our model, the non-perfect velocity plane and the energy-angular momentum plane to the absence of gas in the modeling. For example, we note that the comparison between the model and observations is still inadequate in detail and still reveals some discrepancies including the velocity offset of the Sgr remnant (at $\Lambda$ = 0), it seems that if gas has a role, it could be in slowing down especially the radial velocity of Sgr before its removal, which is validated by simulations in \citep{WangJL2023}. Moreover, because we do not consider gas the energy are expected to be higher compared to the observation. In fact, \citet{Hammer2021} already identified that the energy-angular momentum relationship can be affected by ram pressure, because it acts as a frictional force that does not conserve both quantities, to be more specific, the gas will decrease the radial velocity and energy so we are physically reasonable here. There are also many arguments for a gas-rich progenitor for Sgr. First, all Irr galaxies with a mass-like Sgr are gas-rich when they orbit at more than 300 kpc from the Milky Way and M31 \citep{Grcevich2009}, second, there is a possible relics of gas associated to Sgr found by \citet{Putman2004}, and third, there are important traces of relatively recent star formation and mutiple-populations, at least one billion year ago, in Sgr core and stream \citep{Bonifacio2004,Siegel2007,deBoer2015,2019ApJ...886...57A}.

Moreover, we find that both Milky Way and Sgr masses are relatively modest in our study when compared to that of \citet{Vasiliev2021} model. About the mass of the Milky Way, as known the Sgr initial masses of our model are smaller than that of \citet{Vasiliev2021} model with $M_{star}$=2$\times10^{8}$M$_{\odot}$ and $M_{halo}$ =36$\times10^{8}$M$_{\odot}$. They claimed that  half of the stars are stripped by the Milky Way, and also pointed out that initial mass of the Sgr could be even much higher to a few $\times10^{10}$M$_{\odot}$ according to cosmological abundance \citep{Jiang2000}. Besides this, the \citet{Law2010} model assumed an Sgr total initial mass of 6.4$\times10^{8}$M$_{\odot}$, which is smaller than that in our modeling, but for which no dark matter halo has been assumed. According to \citet{Ibata2020}, the \citet{Law2010} model matches well the Sagittarius stream properties. It seems that high initial masses for Sgr do not favor the reproduction of the stream, as also noticed by \citet{Ibata2020}. We are aware that current work does not allow to give strong constraints on the Sgr initial mass, and we acknowledge that we have not fully investigated the whole mass ranges for both stars and dark matter. 

So our simulations will be considered as providing a new model or new reflection of the Sgr stream, to improve further the modeling, and especially that of the stream kinematics, we will need to implement gas in the Sgr progenitor, or perhaps, alternatively to consider if a massive LMC may also influence the result, although the LMC mass is still uncertain to the community, etc.

\subsection{Disk interactions}  
Finally, the interactions of the Sgr and the Milky Way will possibly excite, flare, warp, ridge, radial motions, bulk motions, oscillations, star formation and so on for the Galactic disc \citep{Laporte2018, 2019MNRAS.485.3134L,2020MNRAS.492L..61L,Bland-Hawthorn2019, wang2018a, wang2018b,wang2019, wang2020a, wang2020b, wang2020c, wang2023a,Yu2021,Antoja2021,Khoperskov2022,Recio2022,2022ApJ...927..131B,2023AJ....165..110Y, 2023ApJ...943...88L,2019ApJ...877L...7W}, and after a quick investigation during our simulation, we have been able to detect flaring features, which is encouraging us to work more about Sgr and Milky Way Disc topics including the chemo-dynamics.

\section{Conclusions}  \label{Concl}

In this work based on N-body simulations, we have almost reconstructed the core \citep{wang2022}, 3D stream, the bifurcations signals in both the leading and the trailing arm, the proper motions and line-of sight velocity variations with $\Lambda$ using one same model, implying that our model is reproducing most properties in the 6D space-phase volume in a qualitative point of view. It represents a good start for our group to towards better understanding for all these features of the Sgr stream. We also notice that our model is different than many previous works to the rotating nature of the Sgr progenitor, and to the relatively small masses for both the Milky way and the Sgr progenitor. 

Our modelling has adopted to relatively modest mass for the Milky Way total mass (5.2$\times10^{11}$ M$_{\odot}$) that matches well its observational rotation curve \citep[see references therein]{Jiao2021, wang2023a} robustly established using Gaia DR2 data \citep{Eilers2019}. The same applies for the ``scaling down''  Sgr total mass (9.3$\times10^{8}$ M$_{\odot}$) with the halo scale radius is 1.6 kpc, which has been modeled as a disc system, together with a relatively massive halo. The variety of masses from different models (\citep{Law2010,Ibata2001,Vasiliev2021} and reference therein) indicates that we are far from being able to predict mass using the present modelling of the Sgr stream. For our modelling, if one would like to consider gas or adopt large scale radius of the dark matter halo, it could naturally be to get a bit higher initial mass, but the change will affect the MW components and initial orbits and infall time.

Furthermore, we have also pointed out some limitations of such a modelling by not being able to reproduce the stream arm behaviors in the radial velocity plane, and even though some quantitative comparisons for 6D properties are also not exact enough for the moment. It may be caused by the fact that our models of the Sgr stream have neglected the gas, while this component might initially prevails in mass when compared to stars. As a first step to towards better understanding of the Sgr system, we currently have used one same model to reproduce 6D properties qualitatively. Although it is still not perfect, the model is very encouraging and promising to do more about fine tuning modelling process, detailed simulations and more comparisons with deeper large sky spectroscopic surveys. 

As the last word, we must admit know different models have different strengths and weaknesses, we are still far from the ultimate truth of the MW and Sgr, more efforts are needed to the community.   


\section*{Acknowledgements}
We appreciate the support of the International Research Program Tianguan, which is an agreement between the CNRS in France, NAOC, IHEP, and the Yunnan University in China. YY thanks the National Natural Foundation of China (NSFC No. 11973042). Simulations in this work were performed at the  High-performent calculation (HPC)  resources MesoPSL financed by the project Equip@Meso (reference ANR-10-EQPX-29-01) of the program "Investissements d'Avenir" supervised by the 'Agence Nationale de la Recherche'. The author gratefully acknowledges the support of CNRS postdoctoral Funding. JLW acknowledge the spport from the China Manned Space Project. We are very grateful to Gary Mamon, Piercarlo Bonifacio,  Eugene Vasiliev, Jiang Chang, Yongjun Jiao, Pierre-Antoine Oria, Rodrigo Ibata, Benoit Famaey and Marcel Pawlowski for the helpful discussions we have had on the Sgr simulation and progress. We thank the referee for giving us precious advices and suggestions that have seriously improved the manuscript. We are grateful to Phil Hopkins who kindly shared with us the access to the Gizmo code (http://www.tapir.caltech.edu/~phopkins/Site/GIZMO.html).

\section*{DATA AVAILABILITY}
The data underlying this article will be shared on reasonable request to the corresponding author.



\vspace*{-\baselineskip}
\bibliographystyle{mnras}









\bsp	
\label{lastpage}
\end{document}